\documentclass[aps,prb,twocolumn,a4paper,amssymb,showkeys,showpacs,floatfix]{revtex4-1}
\usepackage{graphicx}
\usepackage{epsf}
\usepackage{dcolumn}
\usepackage{bm}
\usepackage{amsmath}
\usepackage{amssymb}
\usepackage{color} 
\usepackage{ulem}

\begin{document}

\title{A CDMFT study of antiferromagnetic transition in the square-lattice 
Hubbard model: \\ optical conductivity and electronic structure}
\author{Toshihiro Sato$^1$}
\author{Hirokazu Tsunetsugu$^2$}

\affiliation{
$^1$Computational Condensed Matter Physics Laboratory, 
RIKEN, Wako, Saitama 351-0198, Japan\\
$^2$The Institute for Solid State Physics, The University of Tokyo, 
Kashiwa, Chiba 277-8581, Japan}
\date{\today}

\begin{abstract}
We numerically study optical conductivity $\sigma (\omega )$ 
near the ``antiferromagnetic''  
phase transition in the square-lattice Hubbard model at half filling.
We use a cluster dynamical mean field theory 
and calculate conductivity including vertex corrections, 
and to this end, we have reformulated 
the vertex corrections in the antiferromagnetic phase.  
We find that the vertex corrections change various important 
details in temperature- and $\omega$-dependencies of conductivity 
in the square lattice, and this contrasts sharply the case of 
the Mott transition in the frustrated triangular lattice.  
Generally, the vertex corrections enhance variations in 
the $\omega$-dependence, and sharpen the Drude peak and a high-$\omega$ 
incoherent peak in the paramagnetic phase.   
They also enhance the dip in $\sigma (\omega )$ at $\omega$=0 
in the antiferromagnetic phase.   
Therefore, the dc conductivity is enhanced 
in the paramagnetic phase and suppressed in the antiferromagnetic 
phase, but this change occurs slightly below the transition temperature.  
We also find a temperature region above 
the transition temperature in which the dc conductivity 
shows an insulating behavior but $\sigma (\omega )$ retains 
the Drude peak, and this region is stabilized by the vertex corrections. 
We also investigate which fluctuations are important in the vertex 
corrections and analyze momentum dependence of the vertex function in detail.  
\end{abstract}

\maketitle

\section{Introduction}
\label{sec:Int}

Dynamical mean field theory (DMFT) \cite{DMFT-bethe-sum-1} 
has advanced the investigation of many aspects in 
strongly correlated electronic systems described by the Hubbard models.
This powerful approach is exact on the Bethe lattice with 
an infinite coordination number and has been very successful 
in demonstrating the Mott metal-insulator transition induced 
by electron correlations \cite{DMFT-bethe-sum-1}, 
as well as magnetically ordered states 
by capturing short-range correlations.
Cluster extensions of the DMFT 
i.e., dynamical cluster approximation (DCA) \cite{DCA} 
and cluster DMFT (CDMFT) \cite{CDMFT},
which include both on-site and short-range correlations inside the cluster, 
has made steady progress in our understanding of the Mott transition
\cite{CDMFT-triangular-para-Mott transition-1,%
CDMFT-square-mag-Mott transition-1,
CDMFT-kagome-para-Mott transition-2,
CDMFT-square-para-Mott transition-2,
CDMFT-triangular-para-Mott transition-3,
CDMFT-square-para-Mott transition-3,
DCA-square-para-Mott transition-4,
CDMFT-triangular-para-Mott transition-4,
CDMFT-triangular-para-Mott transition-5,
CDMFT-triangular-para-Mott transition-6,
CDMFT-kagome-para-Mott transition-7}.  
Examples include the phase diagram in the parameter space of 
temperature and Coulomb repulsion strength 
\cite{CDMFT-square-para-Mott transition-3,%
CDMFT-triangular-para-Mott transition-4,
CDMFT-triangular-para-Mott transition-5}
and thermodynamic criticality of the Mott transition
\cite{DMFT-bethe-para-Mott criticality-1,%
CDMFT-anlsotropic triangular-para-Mott criticality-2,
OC w/ VC-CDMFT-triangular-para-1}.
A typical realization of the square-lattice Hubbard model is 
cuprate superconductors and related materials. 
Pseudogap state and superconductivity 
in those systems have been studied actively by using both 
the DCA 
\cite{DCA-frustrated square-mag-pseudogap-1,%
DCA-frustrated square-para-pseudogap-superconductivity-2,
DCA-frustrated square-para-pseudogap-3}
and CDMFT approaches
\cite{CDMFT-frustrated square-para-pseudogap-1,%
CDMFT-frustrated square-para-pseudogap-2,
CDMFT-square-mag-pseudogap-3,
CDMFT-frustrated square-para-pseudogap-4,
CDMFT-frustrated square-para-pseudogap-superconductivity-5,
CDMFT-frustrated square-para-pseudogap-superconductivity-6,
CDMFT-square-para-pseudogap-superconductivity-7,
CDMFT-frustrated-square-para-seudogap-superconductivity-8}.
Another typical realization of the two-dimensional Hubbard model is 
the organic materials $\kappa$-(BEDT-TTF)$_{2}$X, and this has a triangular 
lattice structure.  
The CDMFT calculation has demonstrated a reentrant behavior 
of the Mott transition in an anisotropic triangular 
lattice \cite{CDMFT-triangular-para-Mott transition-4}, 
which is consistent with experimental results 
in some members of this material.

Among studies for advancing our understanding of physical properties 
in the strongly correlated electronic systems, 
transport properties are an active topic of research 
and several experimental results have been reported. 
As a typical example, optical conductivity 
provides useful information on charge dynamics,
in particular, effective mass, transport scattering process 
as well as electric structure. 
In the previous theoretical works of the DMFT 
\cite{OC-DMFT-bethe-para-1,%
OC-DMFT-bethe-para-2,
OC-DMFT+LDA-1,
OC-DMFT-bethe-para-3,
OC-DMFT-bethe-para-4,
OC-DMFT-frustrated square-1,
OC-DMFT-twoorbital triangular-1}
and the CDMFT,  
\cite{OC w/o VC-CDMFT-square-1,%
OC w/o VC-CDMFT-square-2}
optical conductivity of the Hubbard model has been calculated 
simply by convoluting single-electron Green's functions.
These calculations have captured a clear difference in 
charge dynamics between the metallic and the insulating states.
However, to take into account correlation effects further, 
we need a numerical approach that incorporates nonlocal correlations 
in conductivity beyond the standard formulation.  
This has been put forward by including vertex corrections 
inside the cluster based on the developed cluster extensions
by the DCA \cite{OC w/ VC-DCA-square-para-1}.
They were employed only for the paramagnetic phase in a square-lattice system, 
and the results suggested that the vertex corrections makes a significant 
contribution, in particular at and near half filling.
Our previous study also reported the achievement of the vertex 
correction implementation in the CDMFT for optical conductivity 
\cite{OC w/ VC-CDMFT-triangular-para-1}, 
and it focused on the paramagnetic phase in a triangular-lattice system.
In this case, the effects of the vertex corrections are not drastic, 
and this may be attributed to the weak momentum dependence of 
spin correlations due to frustration in the triangular lattice.

These developments in the two methods are crucial to investigate 
correlation effects on electronic transport,
particularly concerning the Mott transition.
However, it is highly desirable to examine the effects of magnetic 
fluctuation on electronic transport for the case when their divergence 
drives a phase transition.  
This is the main issue of this paper and we are going to 
study optical conductivity near the antiferromagnetic phase transition 
in a square-lattice Hubbard model at half filling. 
%
%
For this purpose, we develop a numerical method for optical conductivity 
including vertex corrections in the CDMFT, 
which include effects of magnetic fluctuations. 
We then use this method and investigate the effect of vertex corrections 
near the antiferromagnetic transition temperature 
in the square-lattice Hubbard model at half filling.  
Note that precisely speaking the antiferromagnetic transition does not occur 
at any finite temperature in an isolated layer of two dimensions, 
but one should understand that our calculations mimics a corresponding study 
on a quasi-two-dimensional system.  

This paper is organized as follows.  
We start in Sec.~\ref{sec:MM} describing our model, and then 
explain a new formulation of vertex corrections in optical conductivity 
developed for the antiferromagnetic phase.  
Before showing results of conductivity, 
we briefly discuss in Sec.~\ref{sec:EP} 
magnetic ordering and change in electronic structure with the ordering.  
In Sec.~\ref{sec:TP}, we show the results of optical conductivity 
including the vertex corrections, and then discuss the effects 
of the vertex corrections on dc conductivity. 
The effects on $\omega$-dependence are discussed in detail 
in Sec.~\ref{sec:VC}, and we also analyze which type of fluctuations 
are important.   
In Sec.~\ref{sec:VC-MD}, we analyze momentum dependence of the 
vertex function and investigate how the dependence changes with temperature. 
Section~\ref{sec:SD} concludes this paper with an extended summary. 

\section{Model and Method}
\label{sec:MM}

\begin{figure*}
\centering
\centerline{\includegraphics[width=0.90\hsize]{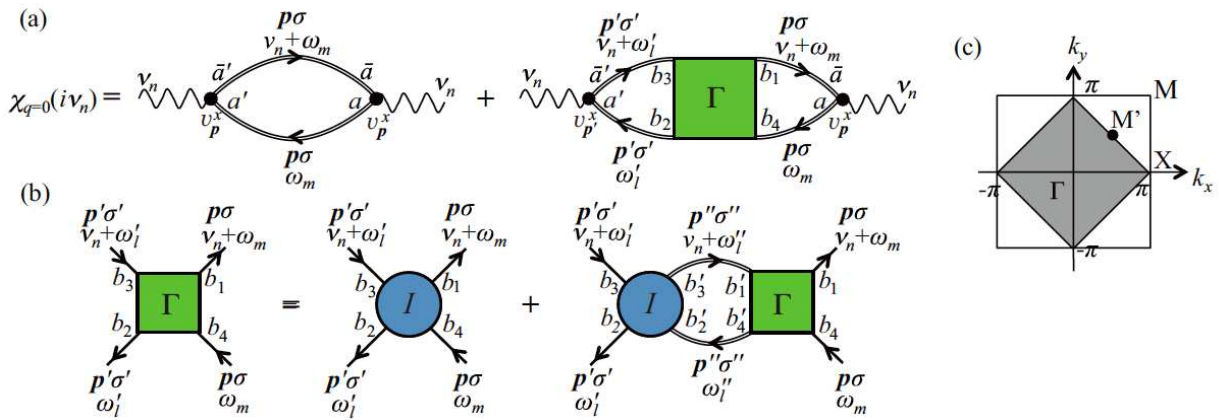}}
\caption{(Color online) Feynman diagrams in the antiferromagnetic phase 
for 
(a) current correlation function $\chi(i\nu_{n})$ in 
Eqs.~(\ref{eq:chi1})-(\ref{eq:OC-1}) and 
(b) full vertex function $\Gamma_{\mathbf{p} \sigma \mathbf{p}' \sigma'}^{b_4b_1b_2b_3}(i\nu_{n})$ in Eq.~(\ref{eq:OC-3mg}).
Results for the paramagnetic phase are obtained 
by omitting the sublattice indices and changing 
the momentum sums to over the original Brillouin zone. 
(c) Brillouin zone of the square lattice. 
Gray zone is the reduced
Brillouin zone in the antiferromagnetic phase.  
}
\label{fig:diag-mg}
\end{figure*}

The model we consider in this paper is a single-band Hubbard Hamiltonian 
on a square lattice at half filling,
\begin{equation}
  H=-t\sum_{\langle i,j \rangle,\sigma}c_{i\sigma}^\dagger c_{j\sigma}
    +U\sum_{i}n_{i\uparrow}n_{i\downarrow}
    -\mu\sum_{i,\sigma} c_{i\sigma}^\dagger c_{i\sigma}.
\label{eq:H-1}
\end{equation}
Here, $t$ is the nearest-neighbor hopping amplitude, 
and $U$ is the on-site Coulomb repulsion. 
The chemical potential $\mu$ is set to $U/2$ to tune the electron 
density at half filling 
$\langle n_{i\uparrow} + n_{i\downarrow} \rangle=1$.  
$c_{i\sigma}$ is the electron annihilation operator at site $i$ 
with spin $\sigma = \uparrow, \downarrow$, 
and $n_{i\sigma} \equiv c_{i\sigma}^\dagger c_{i\sigma}$.
Throughout this paper, the energy unit is $t=1$ and $U$ and $T$ are 
measured in this unit, and all the data are calculated for $U$=6.5.

To take into account both strong short-range electronic correlations and 
magnetic fluctuations, we use the cluster dynamical mean field theory 
(CDMFT)\cite{CDMFT} employing a four-site square cluster.
We compute the single- and two-electron Green's functions inside this cluster
by using the continuous-time quantum Monte Carlo (CTQMC) method based 
on the strong coupling expansion~\cite{CTQMC}.
In this paper, we show the results solely for $U=6.5$.
This choice is close to the value at the critical end point of 
the line of the first-order Mott metal-insulator transition 
in the $U$-$T$ phase diagram determined by 
CDMFT approaches under the condition that no magnetic transitions occur 
\cite{CDMFT-square-para-Mott transition-3}.
Since the square lattice is bipartite, the ground state 
at half-filling electron density has an antiferromagnetic 
order for any $U>0$. 
The calculation by Kent et al.~ showed that the antiferromagnetic 
transition temperature is highest at $U/(12t) \sim 10/12=0.83$ 
in the three-dimensional cubic lattice, where $12t$ is 
the band width.  
In the square lattice, the band width is $8t$ and this 
ratio corresponds to $U/t \sim 0.83 \times 8 \sim 6.7$. 
As this is close to our choice, 
we may expect a high transition temperature, and this is 
an advantage in numerical computation.  
However, we should note that this finite-temperature transition 
is an artifact of the use of the CDMFT, as Mermin-Wagner 
theorem\cite{Mermin} proves its absence in two dimensional lattices.  
Thus, our results for the square lattice 
should be understood as for a corresponding 
quasi-two-dimensional model, in which 
the magnetic order is stabilized by interlayer couplings 
but other physical properties are essentially 
determined in each layer.  
Despite this limitation, the CDMFT approach can take into account 
important short-range quantum and thermal fluctuations 
in our investigation of conductivity near a real 
antiferromagnetic transition.  

In our CDMFT calculations, 
we calculate the single-particle spectrum 
$A_{\mathbf{k} \sigma}(\omega)$ for electron with 
the wave vector $\mathbf{k}$ and spin $\sigma$.  
The dependence on the real frequency $\omega$ is 
obtained by the maximum entropy method (MEM)~\cite{MEM}  
from the Monte Carlo data for imaginary time.  

In the antiferromagnetic phase, we choose the spin axis 
such that local magnetizations point to $\pm z$ direction.
The order parameter is the staggered magnetization 
defined by 
$m_{z}^{(a)}$=%
$\frac{1}{N}\sum_{i \in a}\sum_{\sigma} \sigma \langle n_{i\sigma} \rangle$, 
where the spin is counted as 
$\sigma$=$1(-1)$ for $\uparrow (\downarrow)$, and  
$a$=$A,B$ being the sublattice index.  
$N$ is the total number of sites.  
The relation $m_z^{(B)}$=$-m_z^{(A)}$ holds exactly in the square lattice, 
since the combination of spin inversion and lattice translation 
by (1,0) remains a symmetry operation.  
Because the two sublattices are not equivalent,  
the Brillouin zone halves as shown in 
Fig.~\ref{fig:diag-mg} (c).  
Correspondingly, for labeling electron, one needs an additional 
sublattice index as well as the momentum $(a, \mathbf{p})$ 
in the reduced half Brillouin zone: 
\begin{equation}
c_{A\mathbf{p}} = \frac{c_{\mathbf{p}}+c_{\mathbf{p}+\mathbf{Q}}}{\sqrt{2}}, 
\ \ 
c_{B\mathbf{p}} = \frac{c_{\mathbf{p}}-c_{\mathbf{p}+\mathbf{Q}}}{\sqrt{2}}, 
\label{eq:cAcB}
\end{equation}
where $\mathbf{Q}=(\pi,\pi)$ and the spin index is omitted.  
Each electron propagates from one sublattice 
to the other, and therefore the single-electron Green's function $G$ 
is now a $2\times2$ matrix in the sublattice space. 
Its element $G_{\mathbf{p} \sigma}^{ab}$ denotes the component for the process 
in which an electron is created on the sublattice $b$ and 
then annihilated on the sublattice $a$. 
While the momentum $\mathbf{p}$ is trivially conserved, 
spin projection $\sigma$ conserves because of 
the remaining spin rotation symmetry about the direction 
of the staggered magnetization. 


In two dimensions, optical conductivity $\sigma_{\alpha \alpha'}(\omega)$ 
is a $2\times 2$ matrix, and the linear response theory~\cite{Kubo} 
shows that it is defined by the current correlation function
$\chi_{\alpha \alpha'}(\mathbf q, \omega)$ 
in the limit of the wave vector $\mathbf q \rightarrow \mathbf{0}$. 
Its real part is given in the unit of quantum conductance $(e^2 / \hbar )$ 
as 
\begin{equation}
  \sigma_{\alpha \alpha'}(\omega) 
  = \mathrm{Re} \, 
    \frac{\chi_{\alpha \alpha'}(\mathbf{0},\omega) %
         -\chi_{\alpha \alpha'}(\mathbf{0},0)}%
    {i\omega}    
\label{eq:OC-0}
\end{equation}
where $\mathrm{Re}$ denotes the real part and $\alpha, \alpha' \in \{x,y \}$ 
are the directions of current and electric field, respectively.  
In our case, the conductivity has only an isotropic part, since 
the antiferromagnetic order keeps the lattice rotation symmetry 
about each site, and therefore 
$\sigma_{\alpha \alpha'}(\omega)$%
=$\sigma(\omega)\delta_{\alpha \alpha'}$ and 
$\chi_{\alpha \alpha'}(\mathbf{0},\omega)$%
=$\chi(\mathbf{0},\omega)\delta_{\alpha \alpha'}$ 
not only in the paramagnetic phase but also in the antiferromagnetic phase. 

To take into account correlation effects, 
we have included vertex corrections in CDMFT
in the calculation of $\sigma(\omega)$ 
and reported the results in the paramagnetic phase 
of the frustrated Hubbard model \cite{OC w/ VC-CDMFT-triangular-para-1}.
However, in the antiferromagnetic phase, the Brillouin zone 
halves and this requires the reformulation of the vertex 
corrections, which has not been achieved yet.  
In this paper, we derive this reformulation and 
investigate the effects of magnetic instability 
on transport properties. 
%

Modifying our previous formulation of the vertex corrections, 
we have derived for the antiferromagnetic phase a formula with 
taking account of the new sublattice degrees of freedom.  
The current correlation function is now obtained 
in Matsubara space as\cite{VC}
\begin{eqnarray}
  &&\chi(i\nu_{n})
  =
  \chi_{\mathrm{0}}(i\nu_{n}) 
  +\chi_{\mathrm{vc}}(i\nu_{n}) 
\label{eq:chi1}
\\[4pt]
  &&\chi_{\mathrm{0}}(i\nu_{n}) 
  =
   \sum 
   \bigl( v_{\mathbf{p}}^x \bigr)^2 
   K_{\mathbf{p}\sigma}^{\bar{a} a a' \bar{a'}} (i\nu_{n}; i\omega_{m})
\label{eq:chi0}
\\
  &&\chi_{\mathrm{vc}}(i\nu_{n}) 
  =
  \sum {\sum}'
  v_{\mathbf{p}}^x v_{\mathbf{p}'}^x \, 
  K_{\mathbf{p}\sigma}^{\bar{a} a b_4 b_1} (i\nu_{n}; i\omega_{l}')
  \nonumber \\
  &&\hspace{1.5cm}
  \times
  \Gamma_{\mathbf{p} \sigma \mathbf{p}' \sigma'}^{b_4 b_1 b_2 b_3} (i\nu_{n}) \, 
   K_{\mathbf{p}' \sigma '}^{b_2 b_3 a' \bar{a'}} (i\nu_{n}; i\omega_{m}),
\label{eq:OC-1}
\end{eqnarray}
and the vertex correction $\chi_{\mathrm{vc}}$ is represented 
with the vertex function $\Gamma$.  
Here, $v_{\mathbf{p}}^x = 2t \sin p_x$ is the $x$-component of current, 
and $\sum$ and $\sum '$ are 
$\sum_{\mathbf{p}\sigma}$$\sum_{\omega_{m}}$$\sum_{aa'}$
and 
$\sum_{\mathbf{p}' \sigma'}$$\sum_{\omega_{l}'}$$\sum_{\{ b \} }$, respectively.  
It is important to note that the bare current vertices shown by 
black circles in Fig.~\ref{fig:diag-mg} (a) 
have a special symmetry in the sublattice space. 
Two sublattice indices at each bare vertex should be opposite 
($a$ and $\bar{a}$ etc) and this is because electrons hop 
only between different sublattices in the model (\ref{eq:H-1}).   
We can also show this directly by representing current 
with the new operators, 
\begin{equation}
J_\alpha = \sum_\sigma \sum_{\mathbf{k}}^{\mathrm{full\ BZ}} v_{\mathbf{k}}^\alpha 
 c_{\mathbf{k}\sigma }^\dagger c_{\mathbf{k}\sigma } 
= 
\sum_{\mathbf{p}}^{\mathrm{1/2 \ BZ}} \sum_{a \sigma } v_{\mathbf{p}}^\alpha  
 c_{a \mathbf{p} \sigma }^\dagger c_{\bar{a} \mathbf{p} \sigma}  , 
\label{eq:current}
\end{equation}
where the relation 
$v_{\mathbf{p}+\mathbf{Q}}^\alpha$=$- v_{\mathbf{p}}^\alpha$ is used.   
As depicted in Fig.~\ref{fig:diag-mg} (a), $\chi_0$ corresponds to 
the bubble diagram [the first part on the right hand side (RHS)], and 
$K$ is a product of two single-electron Green's functions shown 
by double lines labeled with sublattice indices at the terminals.  
Note that these single-electron Green's functions include 
the self energy calculated in the CDMFT.  
We directly calculate the single- and two-electron Green's 
functions within the cluster as a function of imaginary time $\tau$ 
by the CTQMC solver.  
The lattice Green's function $G_{\mathbf{p}\sigma}^{ab}(\tau)$ is 
then calculated using the cumulant 
method.\cite{OC w/ VC-CDMFT-triangular-para-1}
Our formulation shown in Eq.~(\ref{eq:OC-1}) and Fig.~\ref{fig:diag-mg} (a) 
uses the approximation for the vertex function $\Gamma$ such that 
its dependence on internal frequencies $\omega_{l}$ and $\omega_{m}'$ 
is averaged over.  

Calculation of the vertex function $\Gamma$ takes a few steps.  
By using the CTQMC method, 
we first calculate directly two-electron Green's functions inside the cluster 
$\mathcal{K}_{i j i' j'}^{\sigma \sigma'}(\tau )$%
$\equiv$%
$\langle c_{i  \sigma}^\dagger (\tau ) c_{j  \sigma} (\tau )
        c_{i' \sigma'}^\dagger(0)     c_{j' \sigma'}(0)     \rangle$, 
where the four sites $i$--$j'$ are all in the cluster.
We then evaluate the irreducible vertex function in the cluster 
$I_{i j i' j'}^{\sigma \sigma'}(i\nu_{n})$ 
by solving the Bethe-Salpeter equation 
$\mathcal{K}_{i j i' j'}^{\sigma \sigma'}(i\nu_{n})$%
$=$%
$K_{i j i' j'}^{\sigma \sigma'}(i\nu_{n})$
$+$%
$\sum_{nm, n' m'} \sum_{\sigma '\!' \sigma '\!'\!'} 
  \mathcal{K}_{i j n m}^{\sigma \sigma'\!'} (i\nu_{n}) \, 
  I_{n m n' m'}^{\sigma'\!' \sigma'\!'\!'}(i\nu_{n}) \, 
  K_{n' m' i' j' }^{\sigma'\!'\!' \sigma'}(i\nu_{n})$,
where $K$ denotes the contribution of a product of 
two single-electron Green's functions~\cite{note2}.
The lattice irreducible vertex is then obtained as 
its Fourier component 
$I_{\mathbf{p} \sigma \mathbf{p}' \sigma'}^{b_4 b_1 b_2 b_3} (i\nu_{n})$%
$=$
$\sum_{ij,i'j'}$%
$  I_{ij,i'j'}^{\sigma \sigma'} (i\nu_{n}) $%
$  e^{i\mathbf{p} \cdot (\mathbf{r}_{i} -\mathbf{r}_{j})
    +i\mathbf{p}' \cdot (\mathbf{r}_{i'}-\mathbf{r}_{j'})}$, 
where the sum is taken over all the combinations under 
the condition that the sites $i$,$j$,$i'$,$j'$ are 
in the sublattice $b_4$,$b_1$,$ b_2$, and $b_3 $, respectively.  
Once $I$ is obtained, 
the lattice reducible vertex $\Gamma$ is calculated by solving 
the Bethe-Salpeter equation that is diagrammatically shown 
in Fig.~\ref{fig:diag-mg} (b),
\begin{eqnarray}
  &&\Gamma_{\mathbf{p} \sigma \mathbf{p}' \sigma'}^{b_4 b_1 b_2 b_3} (i\nu_{n})
  =
  I_{\mathbf{p} \sigma \mathbf{p}' \sigma'}^{b_4 b_1 b_2 b_3} (i\nu_{n})
  +
  \sum 
  \Gamma_{\mathbf{p} \sigma \mathbf{p}'\!' \sigma'\!'}^{b_4 b_1 b_4' b_1'} (i\nu_{n})
  \nonumber \\
  &&\hspace{1.7cm}
  \times
  K_{\mathbf{p}'\!' \sigma'\!'}^{b_4' b_1' b_2' b_3'} (i\nu_{n};i\omega_{l}'\!') \, 
  I_{\mathbf{p}'\!' \sigma '\!' \mathbf{p}' \sigma'}^{b_2'b_3'b_2b_3} (i\nu_{n}).
\label{eq:OC-3mg}
\end{eqnarray}
Here, $\sum$ is a shorthand for 
$\sum_{\mathbf{p}'\!' \sigma'\!'}$$\sum_{\omega_{l}'\!'}$$\sum_{ \{b' \} }$.   
The current correlation in Matsubara space 
$\chi(i\nu_{n})$ is obtained from Eq.~(\ref{eq:OC-1}) with this 
$\Gamma$.  
Finally, we calculate the real part of conductivity $\sigma(\omega)$
by analytic continuation $i\nu_{n}\rightarrow \omega+i0 $ 
by using the maximum entropy method (MEM)~\cite{MEM}.
We note that this new algorithm reproduces our previous formula 
derived for the paramagnetic case~\cite{OC w/ VC-CDMFT-triangular-para-1}
by omitting the sublattice indices and taking wave vector $\mathbf{p}$ 
in the original Brillouin zone as shown in Fig.~\ref{fig:diag-mg} (c).

\section{Magnetic order and electronic properties}
\label{sec:EP}

\begin{figure}
\centering
\centerline{\includegraphics[width=0.8\hsize]{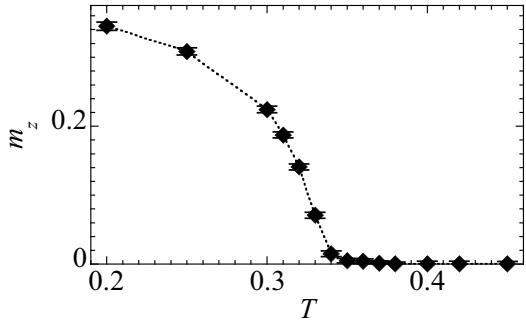}}
\caption{Temperature dependence on staggered magnetization at $U=6.5$.
}
\label{fig:mz}
\end{figure}

\begin{figure}
\centering
\centerline{\includegraphics[width=0.85\hsize]{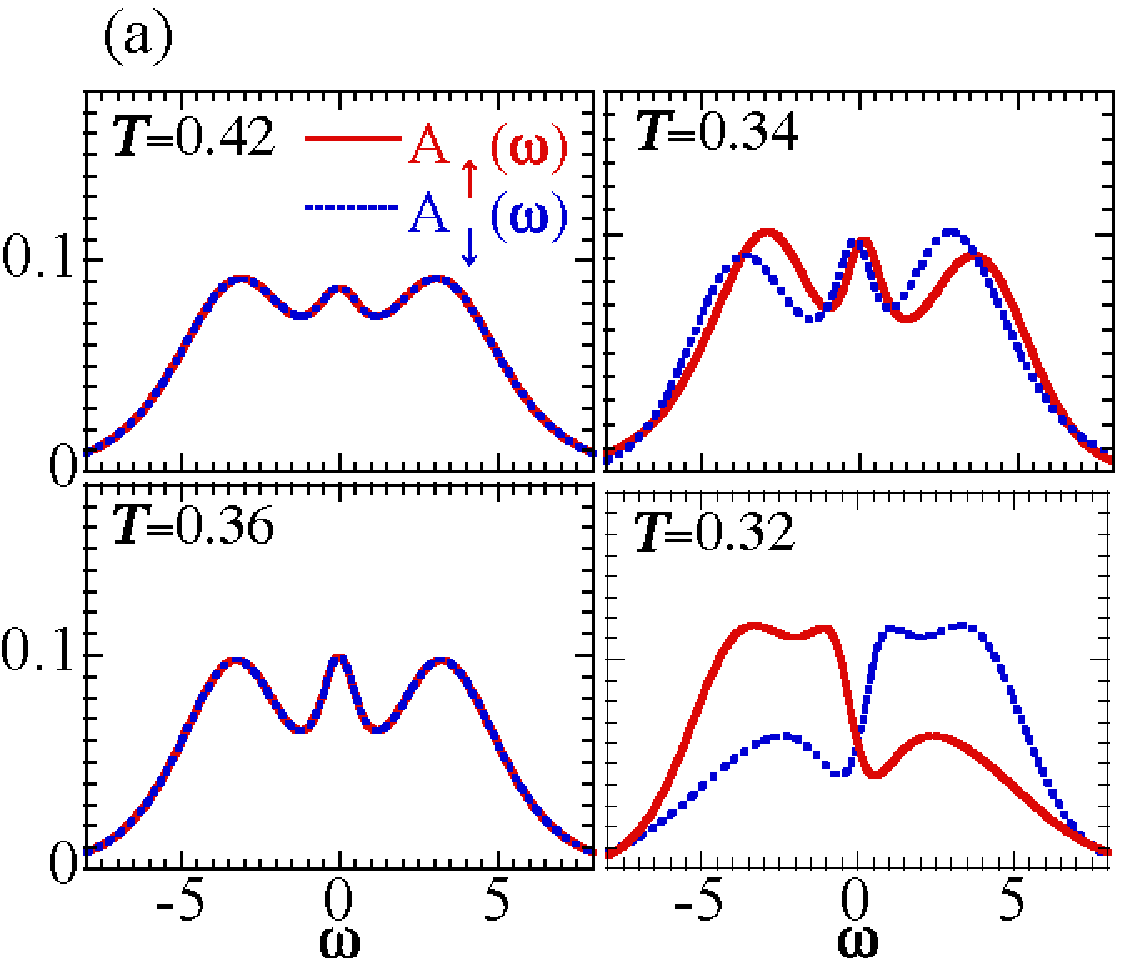}}
\centerline{\includegraphics[width=0.85\hsize]{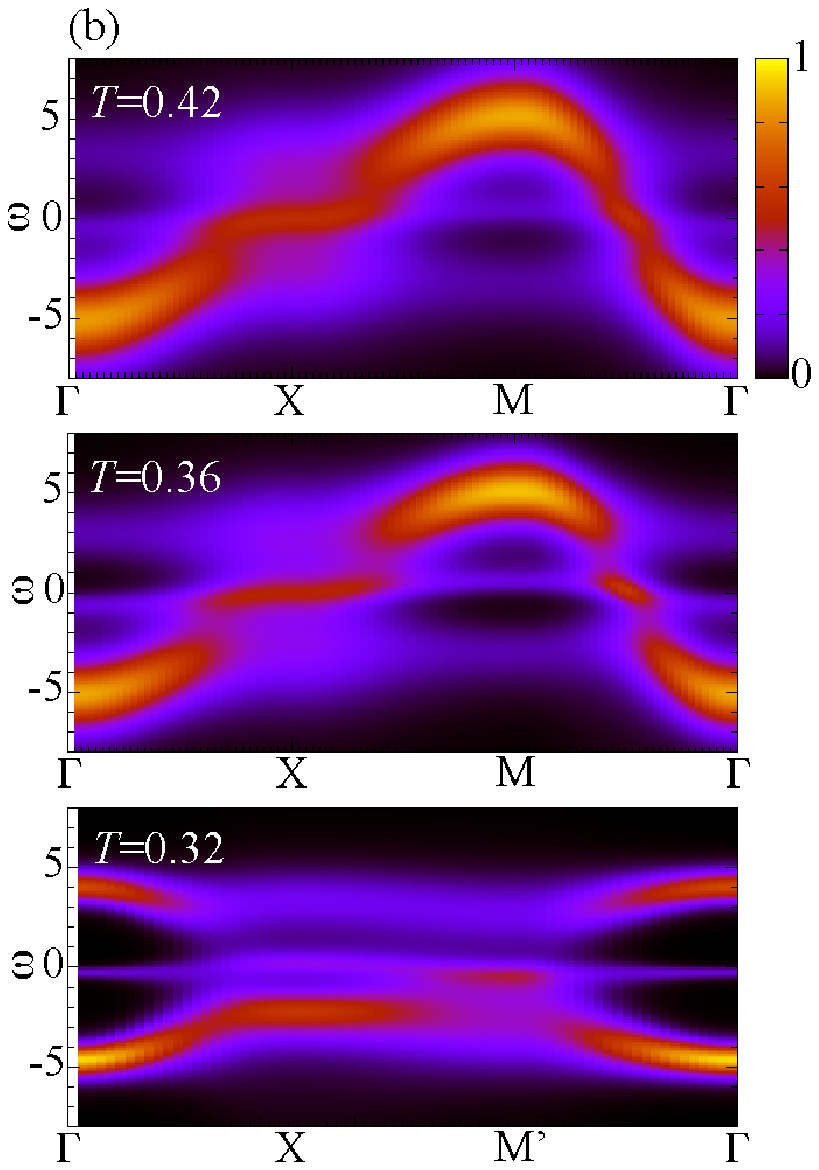}}
\caption{(Color online) 
(a) Local excitations spectrum in the A-sublattice 
$A_{\sigma}^{A}(\omega)$ at four different $T$'s.  
Note that the spectra for different spins are 
degenerate at $T$=0.42 and 0.36.
(b) $\mathbf{k}$-resolved single-particle spectral function 
$A_{\mathbf{k} \uparrow}^{A}(\omega )$ for three different $T$'s
along the path for metallic state and insulating state, 
$\Gamma$-X-M-$\Gamma$ and  $\Gamma$-X-M'-$\Gamma$, respectively
in Fig.~\ref{fig:diag-mg} (c).
The maximum of each spectral function is normalized to be one.
}
\label{fig:mz-A}
\end{figure}

Before discussing conductivity, let us examine in this section 
magnetic order and the variation of electronic state with temperature.  
Figure~\ref{fig:mz}(a) shows the $T$-dependence of the 
staggered magnetization $m_{z} (T)$.
At low temperatures below $T_N$$\sim$$0.34$, 
antiferromagnetic order appears and 
the temperature dependence of is $m_{z}$ well described by 
the mean-field critical exponent of the order parameter 
$\beta = 1/2$. 
This is consistent with the fact that the CDMFT approach is a 
variant of the mean field approximation.  
%

Figure \ref{fig:mz-A}(b) shows the local spectrum of 
single particle excitations in the $A$-sublattice. 
In the paramagnetic phase, it coincides the result in the $B$-sublattice, 
and it is given by averaging the $\mathbf{k}$-dependent 
single-particle spectrum in the whole Brillouin zone, 
$A_{\sigma}^{(A)}(\omega)$=%
$(1/N)\sum_{\mathbf{k}} A_{\mathbf{k}\sigma} (\omega )$, 
with 
$A_{\mathbf{k}\sigma} (\omega )$= %
$-\frac{1}{\pi}{\rm Im}\, {G}_{\mathbf{k} \sigma}(\omega+i0)$.  

In the antiferromagnetic phase, 
the Green's functions contain a spin-dependent component and 
the symmetry of simultaneous spin inversion and sublattice 
exchange leads to the following property
\begin{eqnarray}
&&G_{\mathbf{p} \sigma}^{AA} (i \omega_m ) 
= G_{\mathbf{p} \bar{\sigma}}^{BB}   (i \omega_m )
= g_{\mathbf{p}} (i \omega_m )+ \sigma \Delta_{\mathbf{p}} (i \omega_m ) , 
\nonumber \\
&&G_{\mathbf{p} \sigma}^{a \bar{a}} (i \omega_m )
= G_{\mathbf{p} \bar{\sigma}}^{a \bar{a}} (i \omega_m )
= \tilde{g}_{\mathbf{p}} (i \omega_m ) . 
\label{eq:GF-AF}
\end{eqnarray}
$\Delta$ is the spin-dependent component that appears 
only in the antiferromagnetic phase.  
Due to the sublattice dependencies in Eq.~(\ref{eq:GF-AF}), 
the local spectrum differs between the two sublattices, 
and the value in the $a$-sublattice is given by 
$A_{\sigma}^{(a)} (\omega )$= %
$-\frac{1}{\pi N}{\rm Im}\, \sum_{\mathbf{p}} 
{G}_{\mathbf{p} \sigma}^{aa}(\omega+i0)$.  
The Green's function $G_{\mathbf{k}}$ is now replaced by $G_{\mathbf{p}}^{aa}$ 
and the momentum sum is limited to the reduced Brillouin zone.  
In order to minimize numerical error in $A_{\sigma}(\omega )$, 
we first took $\mathbf{p}$ summation of the imaginary-time 
Green's function and then carried out a transformation to real frequency.  

Variation with $T$ in the local excitation spectrum reproduces 
a known behavior for antiferromagnetic transition in the Hubbard model.  
The spectrum is symmetric in energy, 
$A_{\sigma}(-\omega )$=$A_{\sigma}(\omega )$,  
in the paramagnetic phase, 
and this comes from the particle-hole symmetry due to 
the bipartite lattice structure and the half-filling electron density.  
The spectrum shows three peaks, and this is common in the metallic 
phase of the Hubbard model with large $U$.  
The central peak corresponds to quasiparticle excitations, 
while broad peaks on both sides are the upper 
and lower Hubbard bands~\cite{Fazekas}.    
The central peak sharpens with decreasing temperature 
above $T_N$, implying that 
quasiparticle motion becomes more coherent and the system is metallic.  
We find that there is no indication of pseudogap formation 
and the peak evolution is monotonic down to $T_N$.  
In the antiferromagnetic phase below $T_N$, the excitation spectrum 
splits for different spins, but preserves a generalized 
particle-hole symmetry,  
$A^{(a)}_{\sigma}(\omega )$%
=$A^{(a)}_{\bar{\sigma}}(-\omega )$
=$A^{(\bar{a})}_{\sigma}(-\omega )$.  
This is because the antiferromagnetic ordered state remains 
invariant with respect to the combination of time-reversal 
operation and exchange of the two sublattices.  
Below $T_N$, the spectrum has a dip at $\omega=0$, and this 
deepens with lowering $T$. 
This manifests that the antiferromagnetic phase is insulating.  

%

We discuss the change in electronic structure in more detail by 
examining the momentum resolved single-particle spectrum.  
Figure~\ref{fig:mz-A} (c) presents $A_{\mathbf k \sigma}(\omega)$ 
at three values of $T$.
Figure~\ref{fig:diag-mg} (c) shows color mapping of the spectrum 
in the paramagnetic and antiferromagnetic phases 
along the path $\Gamma$-X-M-$\Gamma$ 
and $\Gamma$-X-M'-$\Gamma$ in the Brillouin zone, respectively.
We only present $A_{\mathbf k \uparrow}(\omega)$ 
because of $A_{\mathbf k \uparrow}(\omega)=A_{\mathbf k \downarrow}(\omega)$
in the paramagnetic state and 
$A_{\mathbf k \uparrow}(\omega)=A_{\mathbf k \downarrow}(-\omega)$ 
in the antiferromagnetic state.
At higher temperature $T$=0.42, in addition to broad peaks 
corresponding to the upper and lower Hubbard bands,
there exists near $\omega$=0 a quasiparticle peak.
With decreasing $T$, the energy dispersion of quasiparticle 
is strongly renormalized to a very flat band, 
implying the strong correlation effects. 
At lower temperature $T=0.32$, this is in the antiferromagnetic 
phase and the low-energy part of the quasiparticle 
band disappears, and an excitation gap opens.
This is due to scatterings by static staggered moment.   
In addition, $A_{\mathbf k \sigma}(\omega)$ now exhibits 
a characteristic peak structure 
near the Fermi energy, which exists near M' point in the Brillouin zone.

\begin{figure}
\centering
\centerline{\includegraphics[width=0.8\hsize]{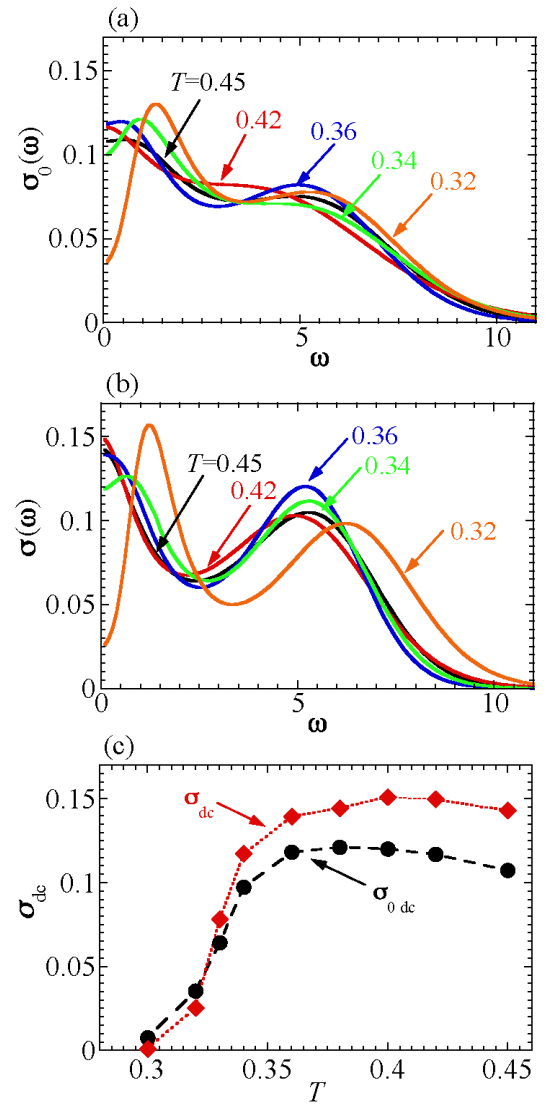}}
\caption{(Color online) 
Optical conductivity (a) without vertex corrections $\sigma_0 (\omega)$ 
and (b) with vertex corrections $\sigma (\omega )$ 
for various $T$'s.
(c) Temperature dependence of dc conductivity with and without 
vertex corrections.  
}
\label{fig:OC-DC-1}
\end{figure}

\section{DC and Optical conductivity}
\label{sec:TP}

Now, we start investigating optical conductivity $\sigma(\omega)$ 
and its dc value.  
A main issue is its variation with temperature 
and the effect of vertex corrections. 
In this section, we are going to investigate characteristics 
in the $T$- and $\omega$-dependence of optical conductivity, 
while examine the vertex corrections in the next section.  

Figure~\ref{fig:OC-DC-1} (a) shows optical conductivity 
before including vertex corrections, $\sigma_0 (\omega)$ . 
They are calculated from $\chi_0$ in Eq.~(\ref{eq:chi0}
at various temperatures both above and below $T_N$. 
The data including the vertex corrections are 
plotted in Fig.~\ref{fig:OC-DC-1}(b). 
First of all, the vertex corrections are noticeable 
and they are large particlarly at low temperatures.  
In our previous result for the frustrated Hubbard 
model on a triangular lattice,\cite{OC w/ VC-CDMFT-triangular-para-1} 
the difference between $\sigma (\omega )$ and $\sigma_0 (\omega )$ 
is quite small.  
Therefore, it is remarkable that the vertex 
corrections are much larger in this unfrustrated system, 
and the corrections are large already in the paramagnetic phase.

One of the most important characteristics is 
the dc conductivity,  
$\sigma_{\mathrm{dc}} = \sigma (\mbox{$\omega$=0})$, 
and this is plotted in Fig.~\ref{fig:OC-DC-1}(c) 
as a function of temperature.  
Values with and without the vertex corrections  
are denoted as $\sigma_{\mathrm{dc}}$ and $\sigma_{\mathrm{0,dc}}$, 
respectively.  
Before investigating $\omega$-dependence, 
we discuss the dc conductivity and the effects of the vertex 
corrections on it.  
When the vertex corrections are not included, 
$\sigma_{0, \mathrm{dc}}$ increases with lowering $T$ 
in the paramagnetic phase, while it decreases 
in the antiferromagnetic phase.  
We find that the dc conductivity shows that the metallic state is 
smoothly connected to the insulating state with varying $T$.
$\sigma_{0,\mathrm{dc}}$ is maximum around $T$$\sim$0.38, 
which is higher than $T_N$. 
The data in Fig.~\ref{fig:OC-DC-1}(c) show that 
the vertex corrections provide opposite contributions 
to $\sigma_{\mathrm{dc}}$ depending on $T$.  
At high temperatures $T$$\ge$0.33, 
which is only slightly below $T_N$$\sim$0.34, 
the correction enlarges the dc conductivity.  
However, the sharp crossover around $T_N$ is enhanced 
to a steeper slope and $\sigma_{\mathrm{dc}}$ is 
suppressed by the vertex correction in the low-temperature 
region.  
It is noticeable that the $\sigma_{\mathrm{dc}}$ maximum 
shifts to a higher temperature $T$$\sim$0.40.

We now examine the $\omega$-dependence of $\sigma (\omega )$.  
A very common feature in all the curves 
in Fig.~\ref{fig:OC-DC-1}(a)-(b) is a broad peak located 
around $\omega \sim U$, and 
this comes from excitations to the upper and
lower Hubbard bands~\cite{Fazekas}.
At higher temperatures $T>0.34$, $\sigma(\omega)$ 
shows a Drude peak around $\omega$=0.  
The system is metallic in this temperature region, 
and the Drude peak comes from motion of quasiparticles.   
The peak is not very sharp but its width gradually narrows 
with decreasing $T$, while its amplitude increases, 
which implies enhancement of coherence in quasiparticle motion.  
The behavior changes below $T_N$.  

With approaching $T_N$, the Drude peak reduces and 
this is attributed to enhanced magnetic fluctuations.
At $T$=0.34, 
the peak that was located around $\omega$=0 now shifts 
to a finite energy $\omega$$\sim$1, and there appears 
a dip at $\omega$=0, which is a characteristic of the insulating phase.  
This comes from the gap opening in the electron 
spectrum discussed for the data in Fig.~\ref{fig:mz-A}.  
With further decreasing temperature down to $T$=0.32, 
the peak shifts towards higher $\omega$ and its intensity grows.  
Correspondingly, the dip at $\omega=0$ deepens.  
As discussed in the previous section, quasiparticle bands 
open a gap in the antiferromagnetic phase. 
The low-energy peak around $\omega$$\sim$1 comes from these 
quasiparticles in the gapped bands.

\section{Vertex corrections in optical conductivity}
\label{sec:VC}

\begin{figure}
\centering
\centerline{\includegraphics[width=0.8\hsize]{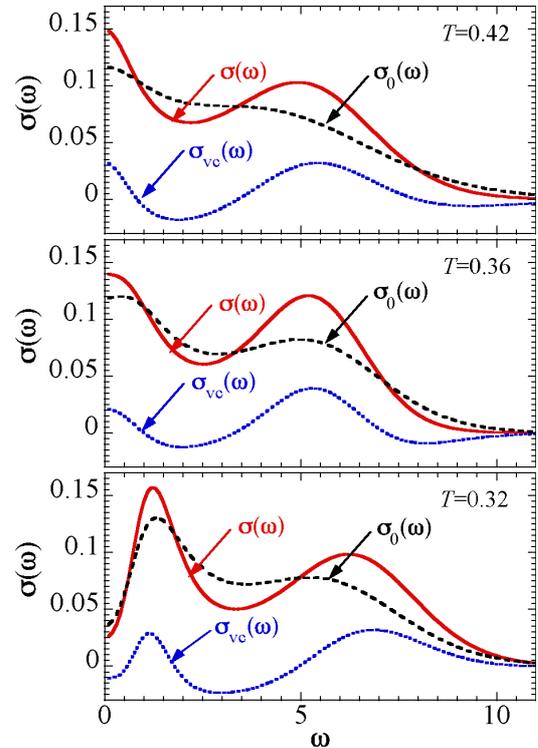}}
\caption{(Color online)
Effects of vertex corrections on optical conductivity 
$\sigma(\omega)$ in the paramagnetic phase ($T$=0.42, 0.36) 
and the antiferromagnetic phase ($T$=0.32).  
$\sigma $=$\sigma_0 $+$\sigma_{\mathrm{vc}} $.  
}
\label{fig:dos}
\end{figure}

The vertex corrections in the dc conductivity were 
studied in the previous section, and 
we now proceed to examine their effects 
on the $\omega$-dependence of the optical conductivity, 
by analyzing the data in Figs.~\ref{fig:OC-DC-1}.   
Detailed comparison is presented in Fig.~\ref{fig:dos} 
for three $T$ values, 
where $\sigma (\omega )$ and $\sigma_{\mathrm{0}} (\omega )$ are the result 
with and without the vertex corrections, respectively.  
$\sigma_{\mathrm{vc}} (\omega )$ is the vertex correction calculated 
from $\chi_{\mathrm{vc}}$.  
General features in their dependence on $\omega$ and $T$ 
are similar to each other, but  
the vertex corrections provide quite pronounced differences 
in the detailed $\omega$ dependence. 
Like in the case of $\sigma_{\mathrm{dc}}(T)$, 
it is interesting that the vertex corrections behave differently  
in between the paramagnetic and antiferromagnetic phases.  

We have found that the vertex corrections generally enlarge 
variations in the $\omega$ dependence of optical conductivity.  
It is also general that the $\omega$-integrated value of the 
correction is positive, and this is consistent with our 
expectation. 
This is because forward scattering processes in quasiparticle 
damping are not effective for current dissipation, 
and the vertex corrections compensate their contributions.  
This also means that the energy gain due to electron kinetic term 
is enhanced.  

Most importantly, the vertex corrections enhance the Drude peak 
and its peak shape becomes noticeably sharper.  
This effect also takes place about the high-energy peak 
around $\omega \sim U$, and 
the broad incoherent peak also becomes noticeably sharper.  
More precisely speaking, 
the part including the vertex corrections provides 
a positive contribution around $\omega =0$ and $\sim U$, 
while negative contribution around $\omega \sim 2$.  
These contributions sharpen the double peak structure 
in the $\omega$-dependence of conductivity.  
The positive contribution around $\omega =0$ is 
gradually suppressed with approaching $T_N$.  

\begin{figure}
\centering
\centerline{\includegraphics[width=0.8\hsize]{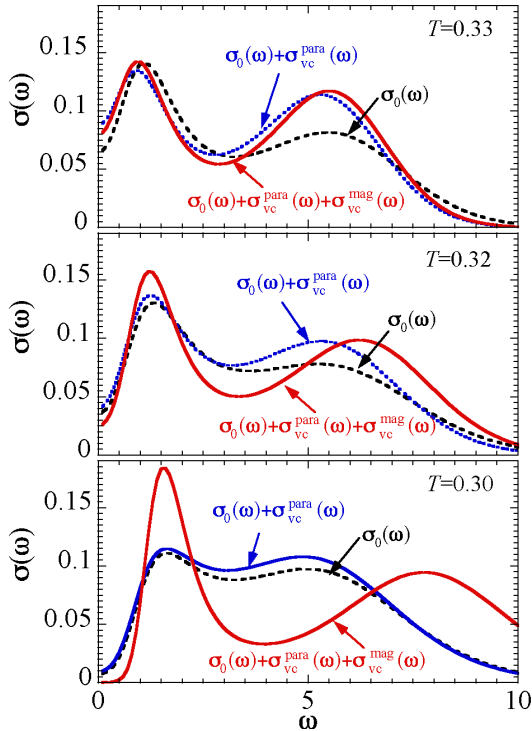}}
\caption{(Color online) 
Effects of the paramagnetic and magnetic parts of 
the vertex corrections to optical conductivity. 
The red curves are the full conductivity $\sigma (\omega )$.  
}  
\label{fig:5-3b}
\end{figure}

We have found the enhancement of $\omega$-dependent structures  
also in the antiferromagnetic phase. 
Most notably this time, the vertex corrections strongly 
suppress conductivity around $\omega=0$, and 
deepen the dip there.  
The low-energy peak around $\omega$$\sim$1 is sharply 
enhanced instead.  
(See also the data in Fig.~\ref{fig:5-3b} (c) at the lower temperature 
$T$=0.30.) 
The peak intensity is enhanced and its width is reduced 
considerably.  
This peak comes from motion of low-energy quasiparticles 
in the gapped bands, and 
this behavior indidates that the vertex corrections strongly 
affects their dynamics.  
The effects in the region of $\omega$$>$2 are similar 
to those in the paramagnetic phase.  
The corrections sharpen the high-$\omega$ peak and deepen the valley.  

Thus, the most striking feature of the vertex corrections is 
their opposite effect in the low-$\omega$ region 
depending on temperature.  
The correction to $\sigma$ value is positive at high temperatures 
and negative at low temperatures, and this sign change 
occurs near $T_N$.  
This behavior was already found in $\sigma_{\mathrm{dc}}(T)$ 
in the previous section, but the $\omega$-dependence 
exhibits this change more clearly.  

An interesting behavior appears in the paramagnetic phase 
near $T_N$.  
With lowering $T$,  as discussed in the previous section, 
$\sigma_{\mathrm{dc}} (T)$ including 
the vertex corrections starts decreasing 
at $T \sim 0.40$, quite higher than $T_N$.  
This is due to antiferromagnetic fluctuations 
strongly enhanced near the transition point.  
However, in this temperature region, 
$\sigma (\omega )$ retains a Drude peak 
and the system remains metallic in this sense.  
This continues down to $T$$\sim$0.36, 
which is close to the temperature of $\sigma_{\mathrm{0,dc}}$ maximum.  
It is interesting that the temperature and $\omega$ dependences 
of conductivity thus behave differently due to the vertex 
corrections in this temperature region.  


Next, we shall examine the effects of antiferromagnetic fluctuations 
in more detail.  
To this end, we separate the vertex correction term in the current 
correlation function into two parts, 
$\chi_{\mathrm{vc}}$=$\chi_{\mathrm{vc}}^{\mathrm{para}}$
+$\chi_{\mathrm{vc}}^{\mathrm{mag}}$.  
The RHS of Eq.~(\ref{eq:OC-1}) is a sum of products of 
four $G$'s, and each $G$ is represented by 
nonmagnetic $g$, $\tilde{g}$ and magnetic $\Delta$ 
depending on sublattice index as shown in Eq.~(\ref{eq:GF-AF}).  
The paramagnetic part $\chi_{\mathrm{vc}}^{\mathrm{para}}$ is 
the sum of the products that do not contain $\Delta$, 
while the magnetic part $\chi_{\mathrm{vc}}^{\mathrm{mag}}$ is 
the sum of all the others.  
Namely, $\chi_{\mathrm{vc}}^{\mathrm{mag}}$ is the part including 
at least one spin-dependent single-electron Green's functions 
among four double lines in the second diagram 
in Fig.~\ref{fig:diag-mg}(a).  

\begin{figure}
\centering
\centerline{\includegraphics[width=0.8\hsize]{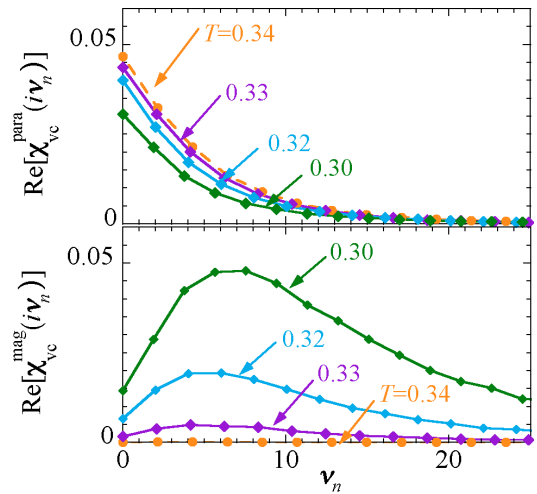}}
\caption{(Color online) 
Contributions of the paramagnetic and magnetic parts  
of the vertex corrections to current correlation function.  
Shown is the dependence on Matsubara frequency $\nu_n$=$2\pi n T$.   
}  
\label{fig:5-3a}
\end{figure}

Figure \ref{fig:5-3a} shows the paramagnetic and magnetic 
parts of the vertex corrections in current correlation function 
at various $T$'s.  
The data are plotted as a function of Matsubara frequency.  
The magnetic part $\chi_{\mathrm{vc}}^{\mathrm{mag}}$ is 
finite only in the antiferromagnetic phase by its definition, 
whereas the paramagnetic part $\chi_{\mathrm{vc}}^{\mathrm{para}}$ 
is sizable in the paramagnetic phase.  
It is remarkable that the two parts behave oppositely to each other 
in the temperature dependence. 
The magnetic part grows very rapidly with lowering temperature. 
In contrast, the paramagnetic part decreases 
slowly but steadily.  
Another difference is about their dependence on Matsubara frequency. 
The maximum of the $\mbox{Re }\chi_{\mathrm{vc}}^{\mathrm{para}}$ stays 
at $\nu_n = 0$ at all $T$'s. 
The magnetic part $\chi_{\mathrm{vc}}^{\mathrm{mag}}$ has 
a maximum at a finite Matsubara frequency and its position 
slowly increases with lowering $T$.  
It is plausible to expect that this peak is related to deepening 
of the conductivity dip at $\omega=0$, and we check this 
by analytic continuation to real frequency $\omega$.  

In Fig.~\ref{fig:5-3b}, $\sigma_{0}$+$\sigma_{\mathrm{vc}}^{\mathrm{para}}$ 
is the optical conductivity calculated from the partial sum 
$ \chi_0 + \chi_{\mathrm{vc}}^{\mathrm{para}}$, 
and it is compared with the full conductivity 
and also the value with no vertex corrections.  
There are two important points.  
The first point is that the magnetic part dominates 
in the vertex corrections 
at low temperature $T=0.32$, although 
the two parts have a similar size at this temperature in 
Fig.~\ref{fig:5-3a}.  
The second point is that the magnetic part changes $\sigma (\omega )$ 
over a much wider range of $\omega$, compared with the paramagnetic part.   
The change due to the paramagnetic part is limited to 
around $\omega \sim U$ and $\omega \sim 0$, and 
the change around $\omega \sim 0$ becomes small at lower temperatures.  
In addition to these two regions, 
the magnetic part also enhances the peak around $\omega \sim 1$ and 
deepens the valley between the two peaks.  

\begin{figure}
\centering
\centerline{\includegraphics[width=0.8\hsize]{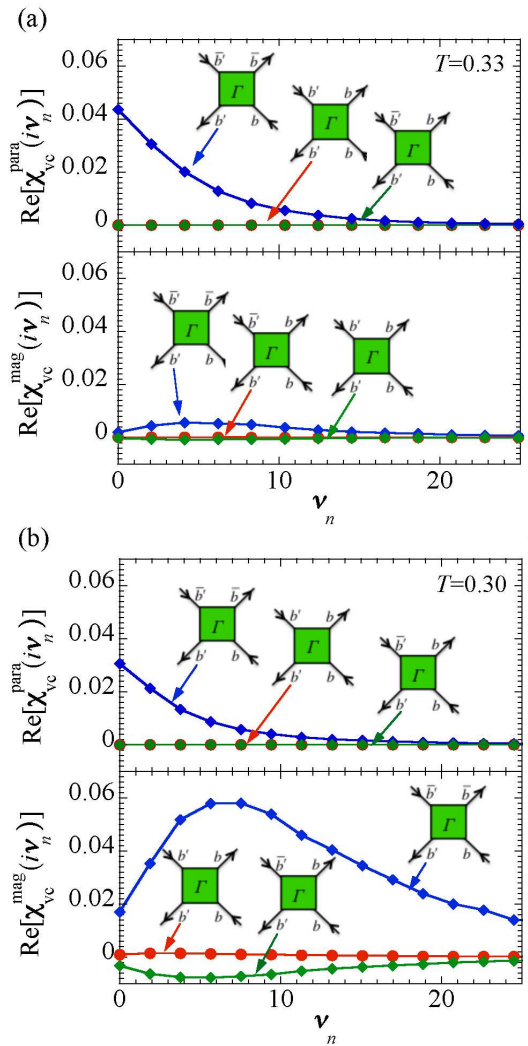}}
\caption{(Color online) 
Comparison of partial $\Gamma$ sums for $\chi_{\mathrm{vc}} (i \nu_{n})$.  
Corresponding to four sublattice indices, 
16 elements of $\Gamma$'s are categorized to the three classes shown.  
Partial sum is taken in Eq. (\ref{eq:OC-1}) for each category. 
Upper and lower panels show the paramagnetic 
and magnetic parts of the vertex corrections.  
Temperature is (a) $T$=0.33 and (b) 0.32.  
}
\label{fig:5-4}
\end{figure}

We now analyze the vertex corrections from a different viewpoint. 
For the full vertex function $\Gamma$, a pair of particle and hole come in 
from external lines 
and they are scattered to another particle-hole pair.  
In the antiferromagnetic phase, particles and holes have also 
sublattice degrees of freedom, and it is interesting to 
examine which combination dominates vertex corrections.  
Corresponding to the four vertices $b_4 b_1 b_2 b_3$ of $\Gamma$ 
in Fig.~\ref{fig:diag-mg}(b), 
there are 16 combinations and 
we group them into three categories. 
Their contributions in $\chi_{\mathrm{vc}}^{\mathrm{para}}$ and 
$\chi_{\mathrm{vc}}^{\mathrm{mag}}$ are shown in Fig.~\ref{fig:5-4}.  
The first category is the ones in which a particle and a hole 
are on the same sublattice on both incoming and outgoing sides 
($b_4$=$b_1$, $b_2$=$b_3$, plotted with red color).  
The second category is the ones in which they are on the opposite 
sublattices on either side ($b_4$=$\bar{b}_1$, $b_2$=$\bar{b}_3$, blue), 
and the third category is the remaining ones (green).  
Each plot shows the sum of the contributions 
from the corresponding combinations.  
The most important point is that the contribution of the second 
category overwhelms the other two in both of 
$\chi_{\mathrm{vc}}^{\mathrm{para}}$ and 
$\chi_{\mathrm{vc}}^{\mathrm{mag}}$, and therefore 
charge and density polarizations made of particles and holes 
on the opposite sublattices play a central role in the 
vertex corrections.  
It is also interesting that the third category has a contribution 
with negative sign.  
However, its amplitude is much smaller than the second category, 
although the amplitude grows at lower temperatures.

\section{Momentum dependence in vertex function $\Gamma$}
\label{sec:VC-MD}

The analysis in the previous section reveals that the most 
important vertex corrections are about the scattering processes 
in which a particle and a hole are on the different sublattices 
to each other both in the initial and final states.  
In this section, we will analyze how this process depends 
on the momentum of particle and hole.  
To this end, we fix the momentum of incoming particle and 
hole at some characteristic $\mathbf{k}$-points 
and examine how they are scattered in the Brillouin zone.  

The square lattice Hubbard model has a Fermi surface with 
rotated square shape at half filling, and 
this is shown in Fig.~\ref{fig:diag-mg}(c).  
It has been well known that the coherence of quasiparticles 
on the Fermi surface strongly depends on their position 
on the surface.   
The functional renormalization-group study\cite{FRG,FRG2} showed that 
quasiparticles near $(\pi, 0)$ or $(0,\pi )$ in the Brillouin zone are 
very incoherent, because they are scattered not only 
by antiferromagnetic spin fluctuations but also by Umklapp processes 
and their renormalized interactions grow rapidly with lowering 
temperature.  
Quasiparticles near $(\pi/2, \pm \pi/2)$ or $(-\pi/2, \pm \pi/2)$ are 
much less incoherent.  
The most incoherent two $\mathbf{k}$-points are called 
\textit{hot spots}, while the least incoherent four are called 
\textit{cold spots}.  
In this section, we also use the terms \textit{hot} and \textit{cold} 
for quasiparticles, if they are at either the hot or cold spots.   

Thus, hot spots and cold spots are opposite limits on the 
Fermi surface, and we set the incoming momentum at 
either of the two.  
As for frequency, we examine the mode with Matsubara 
frequency $\nu_n =0$ as the most characteristic one.  

\begin{figure}
\centering
\centerline{\includegraphics[width=0.95\hsize]{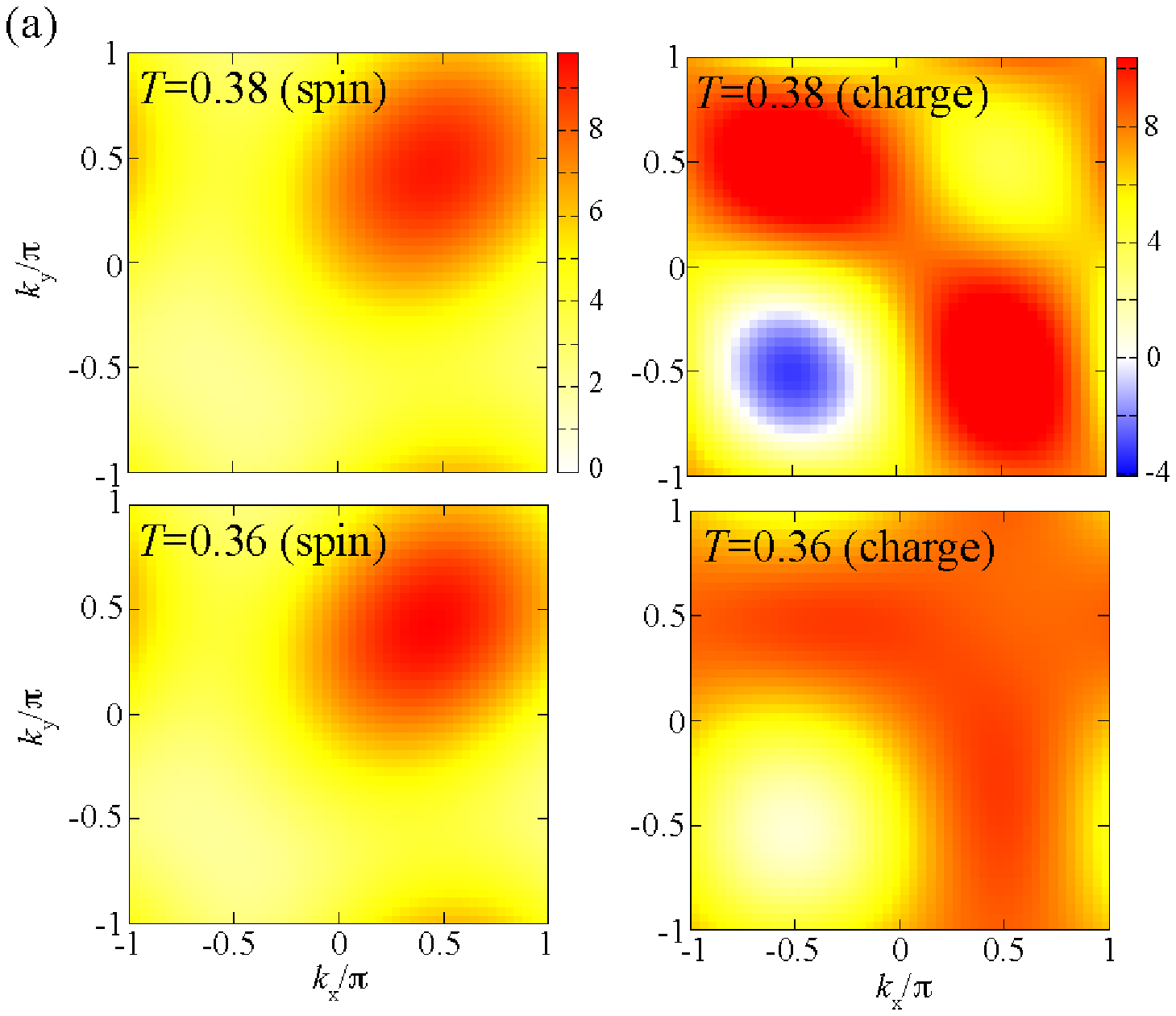}}
\centerline{\includegraphics[width=0.95\hsize]{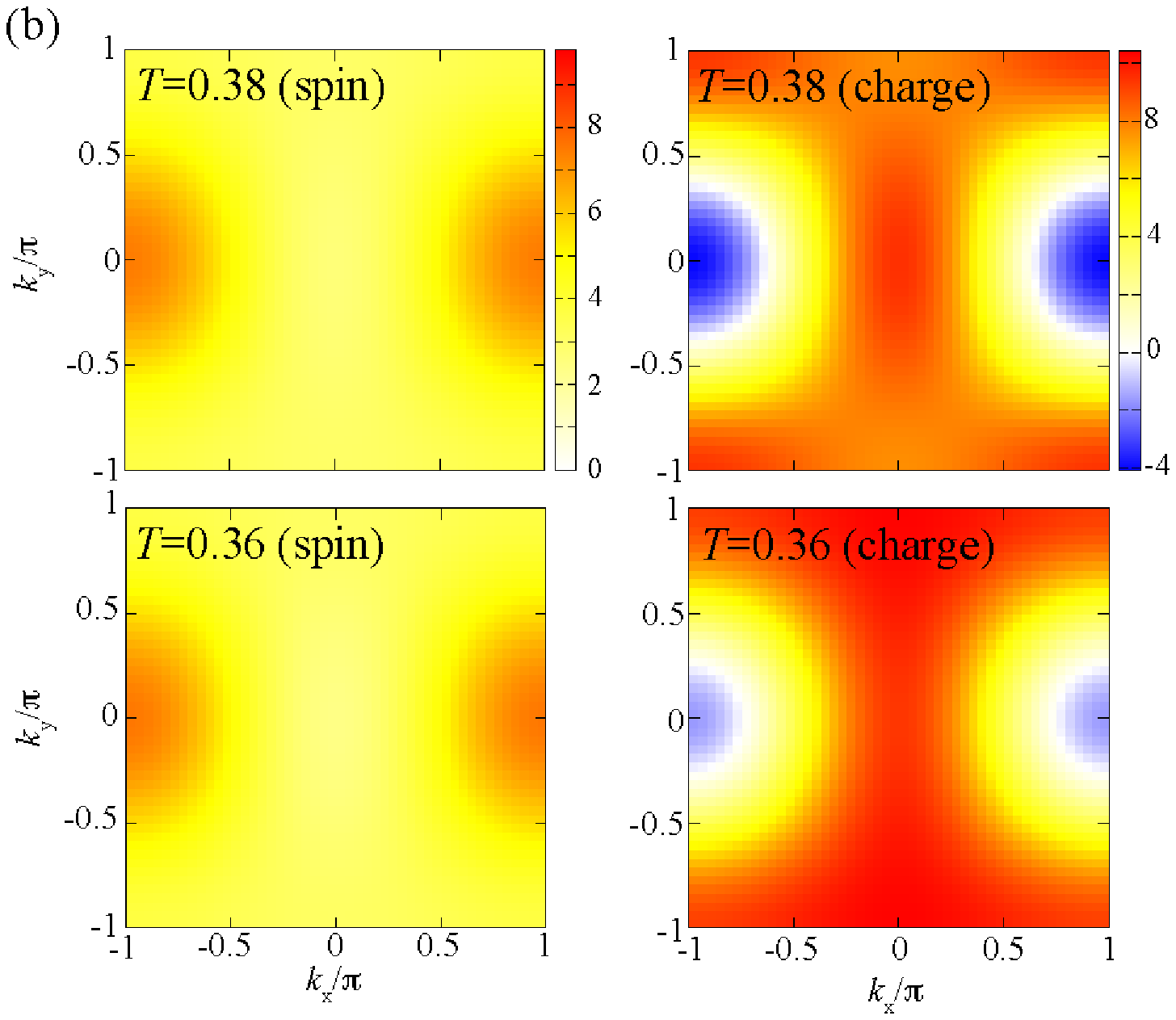}}
\caption{(Color online) 
Momentum-dependence of the spin and charge parts of the full vertex function, 
$\Gamma_{\mathbf k \mathbf k'}^{s,c}(\mbox{$i\nu_{n}$=0})$ 
in Eq.~(\ref{eq:vertex-sc}). 
The case of (a) cold quasiparticles $\mathbf k'=(\pi/2, \pi/2)$ and 
(b) hot quasiparticles $\mathbf k'=(\pi, 0)$.
}  
\label{fig:5-1-2}
\end{figure}

Figure \ref{fig:5-1-2} shows the $\mathbf{k}$-dependence 
of the full vertex function 
$\Gamma_{\mathbf{k} \mathbf{k}'}^{s,c} (\mbox{$i\nu_n$=0})$ 
at $T$=0.38 and 0.36 in the paramagnetic phase.  
The incoming particle is set at the cold spot 
$\mathbf{k}'$=($\pi/2$,$\pi/2$) in (a), 
while 
at the hot spot ($\pi$,0) in (b). 
The incoming hole is at $-\mathbf{k}'$.  
$\Gamma^{c}$ and $\Gamma^{s}$ are the charge and spin vertex 
defined by 
\begin{equation}
\Gamma_{\mathbf{k} \mathbf{k}'}^{c} 
\equiv 
\sum_{\sigma , \sigma' } \Gamma_{\mathbf{k}\sigma \mathbf{k}' \sigma'} ,  
\ \ \ 
\Gamma_{\mathbf{k} \mathbf{k}'}^{s} 
\equiv 
\sum_{\sigma , \sigma' } \sigma \sigma' 
\Gamma_{\mathbf{k}\sigma \mathbf{k}' \sigma'} . 
\label{eq:vertex-sc}
\end{equation}
where all the Matsubara frequencies are $i \nu_n$=0.  
Note that the vertex function $\Gamma$ does not have 
sublattice indices in the paramagnetic phase 
and the details of $\Gamma$ was explained in our previous 
work \cite{OC w/ VC-CDMFT-triangular-para-1}.
The data in Fig.~\ref{fig:5-1-2} provide important information 
on residual effective interactions between quasiparticles.  
Note that optical conductivity is the response of charge current 
and therefore only the charge vertex contributes to $\chi_{\mathrm{vc}}$.  

First, we examine the difference in the $\mathbf{k}$-dependence 
between the spin and charge vertices; more specifically, 
dependence on the momentum difference 
$\Delta \mathbf{k}$$\equiv$$\mathbf{k}-\mathbf{k}'$. 
The most prominent difference is that forward scatterings are 
dominant and have a positive amplitude 
in the spin vertex, while they are very weak 
in the charge vertex.  
Here, forward scatterings refer to the cases of small 
$|\Delta \mathbf{k}|$. 
In the charge vertex, $\Gamma$ is maximum 
for $\Delta \mathbf{k}$$\sim$($\pi$,0) or (0,$\pi$). 
(This depends slightly on temperature, and this will be 
discussed later.)  
These features are common for both incoming quasiparticles 
on the cold spot and those on the hot spot. 
It is also interesting that the forward scatterings 
have a small negative amplitude in the hot spot case.  
However, the $\mathbf{k}$-dependence becomes quite weak 
at lower temperature $T$=0.36 in the charge vertex. 

Secondly, let us compare the behaviors between the cold 
and hot quasiparticles.  
As expected, $\Gamma$ is considerably larger for the hot quasiparticle.   
An interesting point is that the difference is evident only 
in the charge vertex, while the difference in the spin vertex 
is small.  
One difference in the spin vertex is that $\Gamma$ is minimum 
and negative 
for $\Delta \mathbf{k}$$\sim$($\pi$,$\pi$) at the 
cold spot, while the minimum position is 
$\Delta \mathbf{k}$$\sim$($\pi$,0) for the hot spot. 
Otherwise, the difference in the $\mathbf{k}$-dependence 
is rather weak in the spin vertex.  

The difference between cold and hot quasiparticles 
also appears in the evolution with temperature in the charge vertex.  
$\Gamma^c$ decreases at low temperature for the cold quasiparticle 
and strongly increases for the hot quasiparticle, 
which is again consistent 
with the momentum dependence of quasiparticle lifetime.  
Another interesting point related to this is that 
the $\mathbf{k}$-points with large $\Gamma$ for the hot quasiparticle 
distribute very widely in the Brillouin zone. 
These points are around the lines ($k_x$,$\pi$) and (0,$k_y$), 
and the largest $\Gamma$ value is at $\mathbf{k}$=(0,$\pi$).  
Since this is about the charge vertex, this momentum 
difference $\Delta \mathbf{k}$=($\pi$,$\pi$) 
indicates the importance of Umklapp scatterings.

\begin{figure}
\centering
\centerline{\includegraphics[width=0.95\hsize]{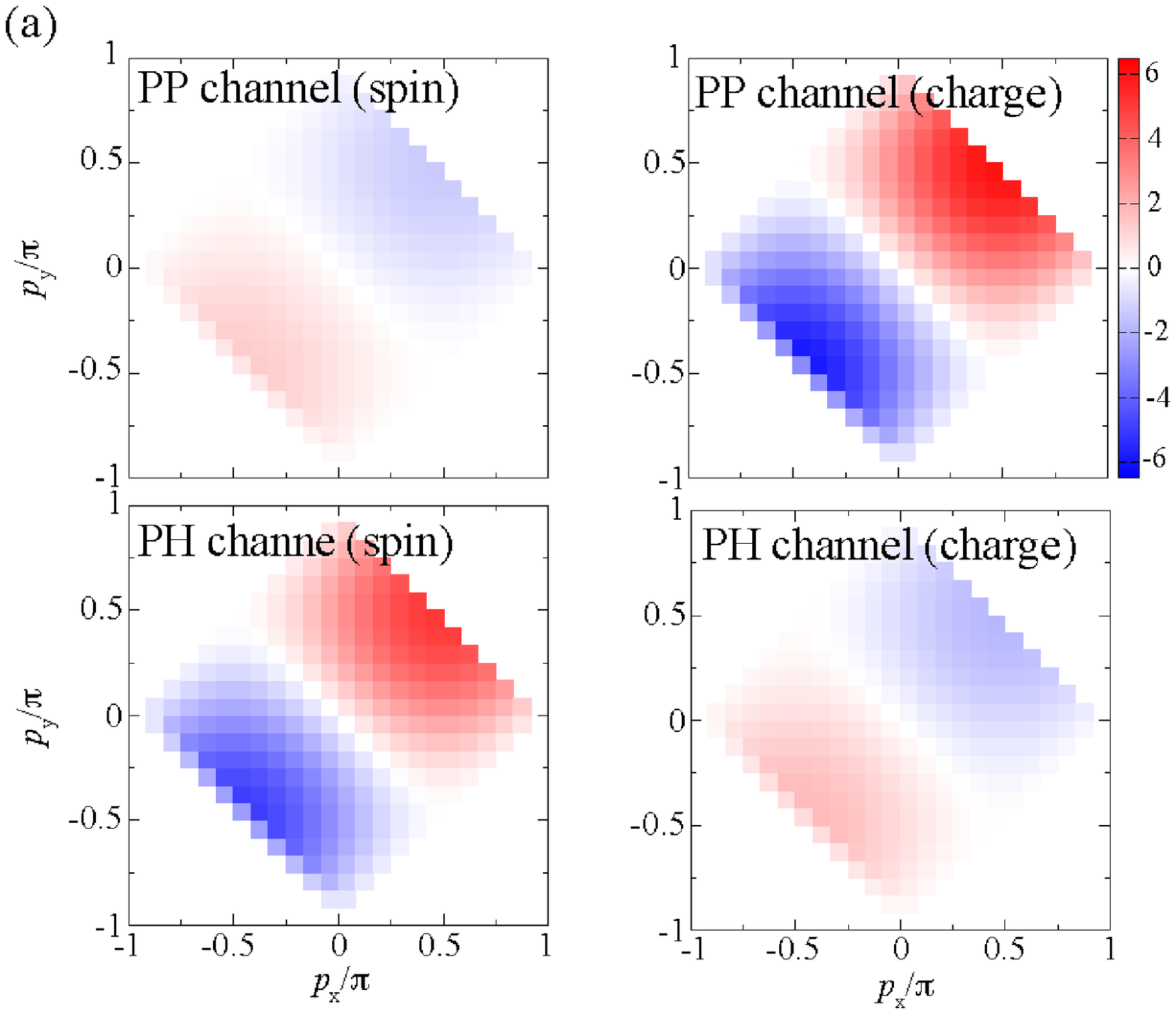}}
\centerline{\includegraphics[width=0.95\hsize]{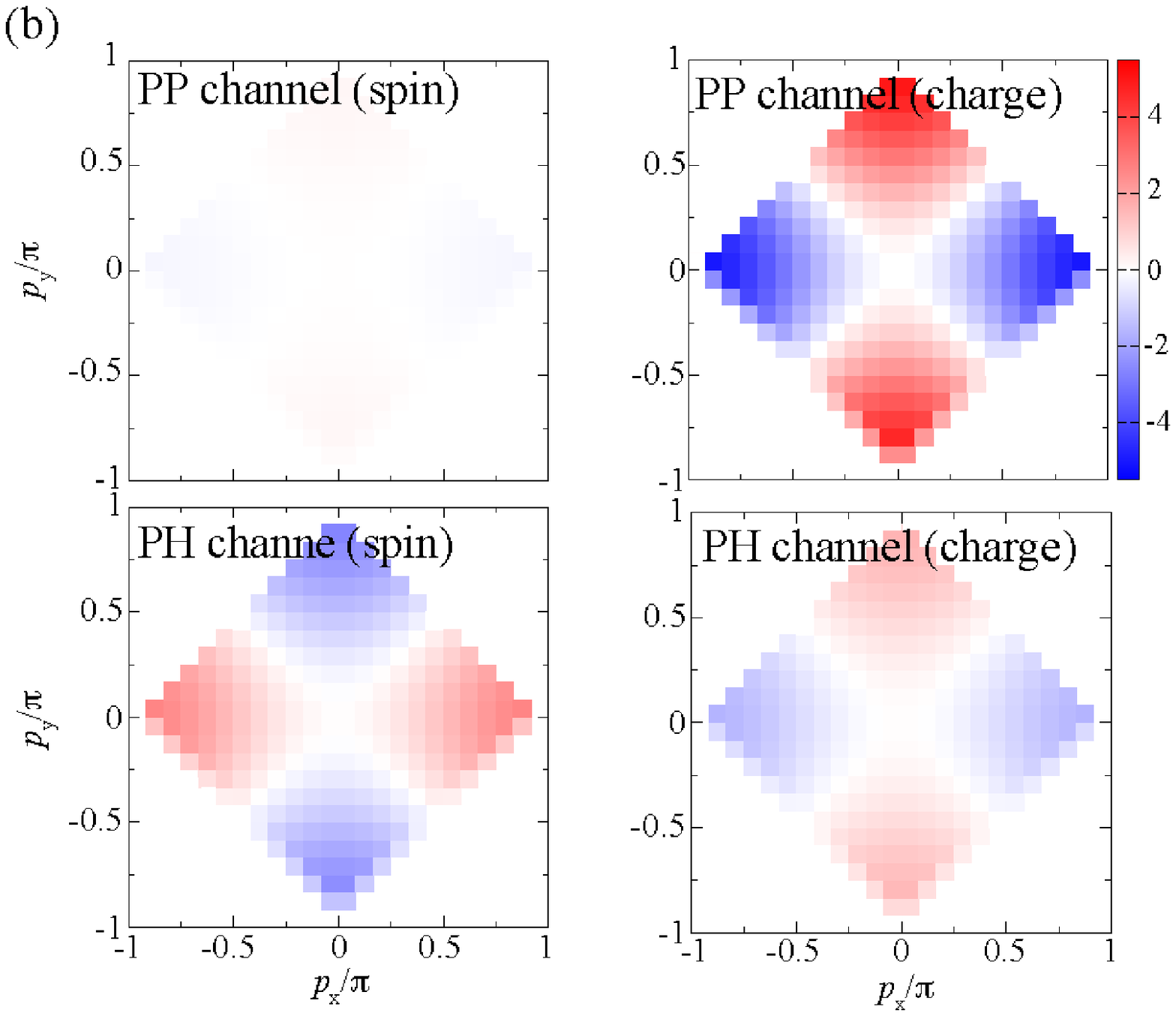}}
\caption{(Color online)
Spin vertex 
$\sum_{\sigma\sigma'} \sigma \sigma' 
\Gamma_{\mathbf{p} \sigma \mathbf{p}' \sigma'}^{b\bar{b}b'\bar{b'}}
(\mbox{$i\nu_{n}$=0})$ 
(left column) and charge vertex 
$\sum_{\sigma\sigma'} 
\Gamma_{\mathbf{p} \sigma \mathbf{p}' \sigma'}^{b\bar{b}b'\bar{b'}}
(\mbox{$i\nu_{n}$=0})$ 
(right column) 
at $T$=0.33.  
The case of 
(a) $\mathbf{p}'$=($\pi$/2, $\pi$/2) and 
(b) $\mathbf{p}'$=($\pi$, 0).
In each case, the particle-particle (PP) channel 
$(b\bar{b}b'\bar{b'})$=(ABAB) and (BABA) 
is shown in the first row, and 
the particle-hole (PH) channel 
$(b\bar{b}b'\bar{b'})$=(ABBA) and (BAAB) 
is shown in the second row.  
}
\label{fig:5-8}
\end{figure}

The full vertex function $\Gamma_{\mathbf{p} \mathbf{p}'}^{s,c}$ 
in the antiferromagnetic phase 
is shown in Fig.~\ref{fig:5-8}.   
The data are calculated at $T$=0.33, which is close to $T_N$.  
We define the charge and spin vertices in the same way as before, 
but one should note that in the antiferromagnetic phase 
the spin vertex also contribute to $\chi_{\mathrm{vc}}$ 
in the current correlation function.  
This is because the single-electron Green's functions contain 
a spin-dependent component $\Delta$, and products of $\Gamma^{s}$ 
and some $\Delta$'s contribute to $\chi_{\mathrm{vc}}$.  
Now, $\Gamma$ has the sublattice indices and specifically  
we examine the part that a particle and a hole are on the different 
sublattices both on the incoming and outgoing sides, since 
the analysis in the previous section showed that this has 
the dominant contribution.  
In the figure, the case that a particle on one sublattice 
is scattered to a particle on the same sublattice (PP channel) 
and the case that it is scattered to a hole on the same 
sublattice (PH channel) are shown separately.  
Recall that the Brillouin zone is reduced to a half in the 
antiferromagnetic phase.  

First of all, the $\mathbf{p}$-dependence in the PP and PH channels 
is very similar to each other, but the amplitude is different. 
The sign is the same for the case of $\mathbf{p}'$=($\pi$,0), 
but opposite for  $\mathbf{p}'$=($\pi$/2,$\pi$/2).  
Generally, the spin vertex has much larger amplitudes in the PH channel, 
while the charge vertex has larger amplitudes in the PP channel but 
the difference is smaller than the case of the spin vertex.  
It is also general that the spin and charge vertices have 
opposite sign for the global phase of $\Gamma$.  

It is very important that the dependence on 
$\Delta \mathbf{p}$$\equiv$$\mathbf{p}$$-$$\mathbf{p}'$ 
is very different between the two $\mathbf{p}'$ cases, and this is true 
for both of spin and charge vertices.  
This feature is distinct from that in the paramagnetic phase.  

We examine the momentum dependence of $\Gamma$ in more detail.  
For $\mathbf{p}'$=($\pi$/2,$\pi$/2), 
the momentum dependence is mainly dominated by 
$\Delta p_x$+$\Delta p_y$ with much smaller dependence on 
$\Delta p_x$$-$$\Delta p_y$.  
This is similar to the dependence of $\Gamma^{c}$ for the cold quasiparticle 
in the paramagnetic phase.
However, the dependence is much more one-dimensional now and 
the direction perpendicular to the initial wave vector $\mathbf{p}'$ 
provides only small correction.  

The dependence is quite different for $\mathbf{p}'$=($\pi$,0).  
In the reduced Brillouin zone, 
the sign of $\Gamma$ is determined by 
$\pm \cos [\frac12 (\mbox{$\Delta p_x$+$\Delta p_y$} )]$%
$\, \cos [\frac12 (\mbox{$\Delta p_x$$-$$\Delta p_y$} )]$%
$\, \propto$ $\pm$$(\cos \Delta p_x$+$\cos \Delta p_y )$, where 
the sign depends on spin or charge part of the vertex.  
This momentum dependence is the one of nearest-neighbor interactions 
in real space, and this result 
indicates that nearest-neighbor correlations are 
dominant for the vertex corrections of those quasiparticles near ($\pi$,0).

\section{Summary}
\label{sec:SD}
In this paper, we have studied optical conductivity near 
the antiferromagnetic transition in a square-lattice Hubbard model 
at half filling.
To calculate optical conductivity, 
we have used a cluster dynamical mean-field approach for 
obtaining single- and two-electron Green's functions.  
For taking account of electron correlation effects, 
we have derived a new formula of the vertex corrections 
in the antiferromagnetic phase based on our previous one for the 
paramagnetic phase.  
We have found that the vertex corrections change various important 
details in temperature and frequency dependence of conductivity 
near the antiferromagnetic transition.  
This point differs from our previous study on optical conductivity 
near the Mott transition in a frustrated triangular lattice. 

An important effect of the vertex corrections is that they 
enhance variations in frequency dependence of conductivity: 
the Drude peak in the paramagnetic phase is enhanced and 
the broad incoherent peak related to the Hubbard band is sharpened.  
The valley in the frequency dependence between these two peaks 
is deepened.  
Optical conductivity shows a dip at $\omega$=0 in the antiferromagnetic 
phase, and the vertex corrections also enhance this dip.  
Another important finding is about a temperature region 
just above the antiferromagnetic transition temperature.  
In this region, dc conductivity decreases with lowering temperature, 
which is similar to a pseudogap phase in the doped case, but 
the electron excitation spectrum shows no indication of 
pseudogap behavior.  
Optical conductivity also have the Drude peak located at $\omega$=0.  
This temperature region exists before including the vertex corrections, 
but the corrections extends the region much wider.  
These are main results directly related to observable 
properties of optical conductivity.  

For better understanding of the vertex corrections, 
we have analyzed which types of fluctuations are important 
in the formula.  
The formula shows that the vertex corrections are determined 
by the vertex function and four single-electron Green's functions.  
Concerning the part of the Green's functions, 
their spin dependent components provide a dominant contribution 
in the antiferromagnetic phase.  
Some of them couple with the spin part of the vertex function, 
while some of the others couple with the charge part, and both 
contribute to conductivity.  
Concerning the part of the vertex function, a predominant 
contribution is given by the scattering processes 
of polarization made of a particle on one sublattice 
and a hole on the other sublattice.  

We have also studied the momentum dependence of the vertex function.  
We have found that the momentum dependence differs 
significantly in the paramagnetic phase 
between the charge vertex and the spin vertex.  
An important point is that for quasiparticles near 
($\pi$,0) or (0,$\pi$) in the Brillouin zone 
the vertex functions are strongly enhanced near the antiferromagnetic 
instability and the dependence on the scattered momentum 
indicates the importance of Umklapp scatterings. 
In the antiferromagnetic phase, the charge vertex and spin vertex 
functions have a similar momentum dependence 
but the sign is opposite.  
The antiferromagnetic phase also has very different momentum 
dependence of the vertex function between quasiparticles 
at different positions in the Brillouin zone. 
For those at ($\pi$,0) or (0,$\pi$), the momentum dependence 
is dominated by nearest-neighbor correlations. 
For those at ($\pi/2$,$\pm \pi/2$), the momentum dependence 
is quite one-dimensional.  
At this moment, it is not clear yet how these exotic correlations 
affect conductivity, but we believe that these detailed data in the 
vertex function and the vertex corrections obtained in the present work 
will provide useful information in future for constructing 
theories for better understanding of conductivity in the Hubbard model.

The authors are grateful to Kazumasa Hattori for valuable discussions 
and his useful comments.  
Numerical computations have been performed with using the facilities at 
Supercomputer Center in ISSP and Information Technology Center, 
both of which are at the University of Tokyo, as well as 
the RIKEN Integrated Cluster of Clusters (RICC) facility. 
This work has been supported by Grant-in-Aid for Scientific 
Research from MEXT, Japan (No.~25400359).

\end{document}